\documentclass[twocolumn,footinbib,aps^]{revtex4-1}
\usepackage{color}
\usepackage{amsfonts,amssymb,amsmath,mathrsfs}
\usepackage{amsbsy}
\usepackage{graphicx}
\usepackage{hyperref}
\usepackage{qcircuit}

\newcommand{\ket}[1]{\left| #1 \right\rangle}

\begin{document}
\title{Robust implementation of quantum gates despite always-on exchange coupling in silicon double quantum dots}
\author{Utkan G\"ung\"ord\"u}
\email{utkan@umbc.edu}
\affiliation{Department of Physics, University of Maryland Baltimore County, Baltimore, MD 21250, USA}

\author{J.~P.~Kestner}
\affiliation{Department of Physics, University of Maryland Baltimore County, Baltimore, MD 21250, USA}

\begin{abstract}
Addressability of spin qubits in a silicon double quantum dot setup in the (1,1) charge configuration relies on having a large difference between the Zeeman splittings of the electrons. When the difference is not sufficiently large, the rotating wave approximation becomes inaccurate.
We consider a device working in this regime, with always-on exchange coupling, and describe how a CZ gate and arbitrary one-qubit gates which are robust against charge noise can be implemented by smoothly pulsing the microwave source, while eliminating the crosstalk. We find that the most significant deviations from the rotating wave approximation, which are analogous to the Bloch-Siegert shift in a two-level system, can be compensated using local virtual gates.
\end{abstract}

\maketitle

\section{Introduction}
Silicon is a promising platform for practical quantum computing.
Average fidelities reaching 99.93\% have recently been reported for one-qubit gates in Si/SiGe \cite*{Yoneda2017b} and 99.96\% in Si/SiO$_2$ \cite*{Veldhorst2014b,Yang2019b}.
However, a universal quantum computer requires entangling operations as well \cite*{Lloyd1995a}, and the infidelities for two-qubit gates remain orders of magnitude higher \cite*{Brunner2011,Veldhorst2014,Watson2017,Zajac2017a,Xue2019}.
Hyperfine coupling with remnant spinful isotopes such as $^{29}$Si and $^{30}$Si is one source of errors, but the concentration of these isotopes can be reduced down to 800ppm \cite*{Yoneda2017b,Yang2019b,Huang2019} or lower \cite*{Abrosimov2017}, essentially eliminating it as a concern. Another significant source of errors is charge noise, which remains a problem \cite*{Culcer2009,Yoneda2017b,Gungordu2019a,Jock2018,Chan2018a,Connors2019}.
Although it can be partially mitigated when operating at a ``sweet spot" where the exchange is insensitive to the leading order effects of the electrostatic fluctuations in gate voltages \cite*{Zajac2017a}, even there the second order fluctuations can still be large, and there can be fluctuations in the tunnel barrier as well \cite*{Huang2018z}.  Thus, schemes to correct charge noise are of great importance for scalable quantum computation in semiconductor devices.

Quantum dot setups in which the spin of each electron is treated as a qubit are particularly attractive. Typical implementations of one-qubit gates require the ability to address one of the qubits without affecting the other one. In setups using electron dipole spin resonance (EDSR) \cite*{Zajac2017a,Watson2017}, this is done by placing a micromagnet such that one of the electrons is closer to the micromagnet than the other one, providing a strong magnetic field gradient. The resulting difference between the Zeeman energy splittings of the electrons, $\delta E_z$, separates their resonant frequencies and allows implementation of fast \cite*{Russ2017a,Vargas2019} or dynamically corrected \cite*{Gungordu2018a} two-qubit gates to suppress the impact of the charge noise on the exchange.  However, the micromagnet can also couple charge noise directly to the spins through the effective spin-orbit interaction, opening a new way for gate errors to enter.
In setups with electron spin resonance (ESR) \cite*{Veldhorst2014,Huang2019}, on the other hand, the spins remain largely isolated from the electrical environment in the absence of coupling. A difference in resonant frequencies then comes only via a difference in the $g$-factors of the electrons. The effective $g$-factors of the electrons strongly depend on the interface hosting the 2D electron gas (2DEG) \cite*{Ruskov2018,Hwang2017,Ferdous2018}, which is an aspect of the fabrication that cannot be controlled perfectly, and although the resulting $\delta E_z$ can be electrically \cite*{Chan2018a} or magnetically \cite*{Tanttu2019} tunable to a certain extent, the maximum accessible value is typically much less than that provided by a micromagnet \cite*{Watson2017}.

In a double quantum dot setup, this causes two issues with controllability.
First, addressability is diminished because the resonant frequency of the electrons are too similar: tuning the frequency of the magnetic control field into the resonance frequency of one of the electrons causes unwanted dynamics on the other electron, which cannot be neglected under a simple rotating wave approximation when using a reasonable driving strength.
Second, whenever the exchange interaction and the driving are turned on simultaneously, this further leads to exchange-induced crosstalk between the two electrons. In the literature, these issues are also known as classical and quantum crosstalk \cite*{Patterson2019}.

In Ref.~\cite*{Gungordu2018a}, we described a robust pulse sequence which implements a CZ gate in silicon double quantum dots while correcting charge noise. However, this scheme relies on having access to high-fidelity one-qubit gates in the $(1,1)$ charge stability region while also having fully controllable exchange coupling. In devices where addressability is poor, or where exchange cannot be turned off within (1,1) region \cite*{Huang2019}, implementation of high-fidelity one-qubit gates is difficult, even in the absence of any stochastic errors such as charge noise.

In this paper, we simultaneously solve both issues of charge noise and addressability in the presence of an always-on exchange coupling and small $\delta E_z$ by deriving shaped pulses that implement a robust CZ gate, as well as robust, arbitrary one-qubit gates. Although our work is motivated by the experiment in Ref. \cite*{Huang2019} in which the exchange coupling in always turned on and constant, our results are applicable to similar devices where the residual exchange between the electrons is nonnegligible or the control over exchange coupling is strongly bandwidth-limited.

This paper is organized as follows. In Section \ref{sec:model}, we describe the effective Hamiltonian for the silicon double quantum dot in the (1,1) charge stability region, and discuss how it algebraically splits into a pair of independent two-level systems when the rotating wave approximation required for addressability holds. In Section \ref{sec:bloch-siegert}, we show that  although errors from the rotating wave approximation are significant, they can be compensated for by using only local $Z$ rotations. In Section \ref{sec:robustness}, we show how the results from Ref. \cite*{Barnes2015} can be used to suppress charge noise in the two-qubit system, followed by Section \ref{sec:gates} where we explicitly construct robust CZ and robust arbitrary one-qubit gates. Section \ref{sec:conclusion} concludes the paper.

\section{Model}
\label{sec:model}

\begin{figure}
	\includegraphics[width=0.8\columnwidth]{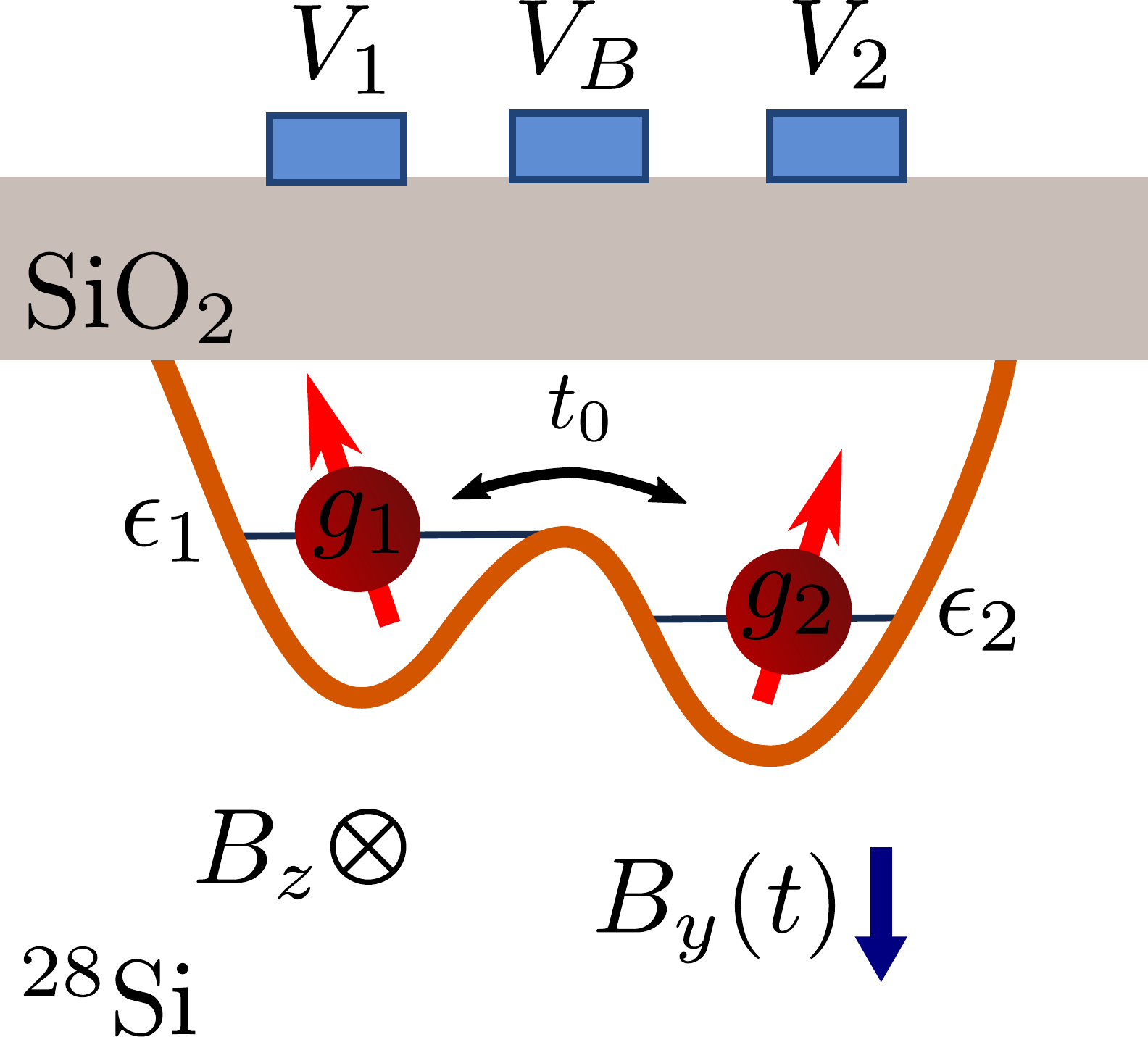}
	\caption{(Color online) Illustration of a gate-defined double quantum dot formed within the 2DEG at the interface of a Si/SiO$_2$ heterostructure, with metal gates on top (some setups may lack the middle gate, which can be used to independently modulate the tunnel barrier to control the energy cost of hopping). Gate voltages are tunable within a range which ensure one electron is loaded in each dot, and they determine the effective $g$-factor of each electron as well as the strength of the effective exchange coupling. In addition to a global magnetic field $B_z$, an oscillatory magnetic field $B_y(t)$ is applied by running a current through a nearby wire (not depicted).}
	\label{fig:setup}
\end{figure}

Electrons in a double quantum dot in the $(1,1)$ charge stability region, as illustrated in Fig.~\ref{fig:setup} can be modeled by the following Hamiltonian \cite*{Yang2011,DasSarma2011}
\begin{align}
H = \begin{pmatrix}
\bar E_z & \frac{{E_\perp^2}^*}{2} & \frac{E_{1,\perp}^*}{2} & 0 & 0 & 0 \\
\frac{E_\perp^2}{2} & \frac{\delta E_z}{2} & 0 & \frac{{E_\perp^1}^*}{2} & t_0 & t_0 \\
\frac{E_\perp^1}{2} & 0 & -\frac{\delta E_z}{2} & \frac{{E_\perp^2}^*}{2} & -t_0 & -t_0 \\
0 & \frac{E_\perp^1}{2} & \frac{E_\perp^2}{2} & - \bar E_z & 0 \\
0 & t_0 & -t_0 & 0 & (U_1-\epsilon) & 0 \\
0 & t_0 & -t_0 & 0 & 0 & (U_2+\epsilon) \\
\end{pmatrix}
\end{align}
in the basis of $\ket{\uparrow \uparrow}, \ket{\uparrow \downarrow}, \ket{\downarrow \uparrow}, \ket{\downarrow \downarrow}, \ket{S(2,0)},\ket{S(0,2)} $, which includes the possibility of tunneling of one of the electrons to its neighboring dot and occupying the singlet states $\ket{S(2,0)}$ and $\ket{S(0,2)}$.
Above, $\bar E_z = (E_z^1 + E_z^2)/2 = \mu_B (g_1 B_z^1 + g_2 B_z^2)/2$ and $\delta E_z = E_z^1 - E_z^2 = \mu_B (g_1 B_z^1 - g_2 B_z^2)$ denote the average and difference in Zeeman energies of the electrons due the strong magnetic field along $z$ provided by an external static magnet, $t_0$ is tunneling energy, $E_\perp^k$ is the Zeeman energy
$i \mu_B g_k B_y(t)$
due to oscillating magnetic field as seen by the $k$th electron, $U_k$ is the on-site charging energy (Coulomb energy due to double occupancy), $\epsilon_k$ is the on-site single occupancy energy of each dot, and $\epsilon = \epsilon_2 - \epsilon_1$ is the ``detuning''. In this setup, both electrons feel the same magnetic field and Zeeman energies are made different via the electrically-tunable difference between effective electron $g$-factors; in particular, there is no magnetic field gradient ($B_z^1 = B_z^2$) due to the lack of a micromagnet.

$E_\perp^k$ is due to the weaker and controllable component of the magnetic field, commonly referred to as the microwave field, which is used for performing single-qubit operations by selectively addressing each spin. To that end, it is modulated at or near the resonant frequency of the target qubit: $E_\perp^k \propto \cos(\omega t)$ with $\omega \sim E_z^k$.  We will thus write \footnote{Any constant drift term is negligible due to the first round of rotating wave approximation that follows below.}
\begin{align}
\Omega_k \cos(\omega t) \equiv \mu_B g_k B_y(t),
\label{eq:g}
\end{align}
where $\Omega_k$ is the slowly varying (compared to the fast modulation at $\omega$) microwave amplitude and $\omega$ is the microwave frequency. When the fast modulation $\hbar\omega$ is tuned to one of the Zeeman splittings $E_z^k$, which are the dominant terms in the effective $4\times 4$ spin Hamiltonian (which will be derived just below Eq.~\eqref{eq:H4x4}) and are typically at least three orders of magnitude larger than the other terms (microwave amplitude $\Omega_k$ and exchange coupling strength), one can use rotating wave approximation to effectively rewrite $E_\perp^k$ as $\Omega_k e^{i \omega t}$ \cite*{Berman2011}; and in what follows, we will always assume that $\omega$ is tuned to be near the resonant frequency of one of the electrons. This particular rotating wave approximation holds very well experimentally.

Furthermore, in the typical experimental regime of $U_k\pm\epsilon \gg t_0, \delta E_z$, one can use a (time-independent) Schrieffer-Wolff transformation to block-diagonalize $H$ and obtain the following spin Hamiltonian \cite*{Russ2017a,Gungordu2018a,Huang2019} \footnote{The smallness factor in the Schrieffer-Wolff transformation is $\approx t_0/(U_k \pm \epsilon) \sim 1/2000$ for a typical experimental value of $t_0 \sim 1$GHz \cite*{Veldhorst2014}, which suppress all perturbative terms with the exception of those due to the $t_0$ term itself (which lead to $J$) in the original spin Hamiltonian $H$.}
\begin{align}
H \approx \begin{pmatrix}
\bar E_z & \frac{{E_\perp^2}^*}{2} & \frac{{E_\perp^1}^*}{2} & 0 \\
\frac{E_\perp^2}{2} & \left(\frac{\delta E_z}{2} - \frac{J}{2}\right) & \frac{J}{2} & \frac{{E_\perp^1}^*}{2} \\
\frac{E_\perp^1}{2} & \frac{J}{2} & \left(-\frac{\delta E_z}{2} - \frac{J}{2}\right) & \frac{{E_\perp^2}^*}{2} \\
0 & \frac{E_\perp^1}{2} & \frac{E_\perp^2}{2} & - \bar E_z
\end{pmatrix},
\label{eq:H4x4}
\end{align}
where $J \equiv 2 t_0^2/(U_1-\epsilon) + 2 t_0^2/(U_2+\epsilon) $ is the tunneling mediated exchange coupling between the two electrons. This Heisenberg-type exchange coupling mixes $\ket{\uparrow\downarrow}$ and $\ket{\downarrow\uparrow}$ states; however, when it is changed slowly (such that $\hbar\dot J/ \delta E_z$ is negligible compared to other terms, assuming $J \ll \delta E_z$ \cite*{Gungordu2018a}) or held fixed (as is the case we will focus on), one can instead work in the eigenbasis of the Hamiltonian (at $\Omega_i=0$), in which all states become decoupled and the exchange coupling becomes Ising-like \cite*{Meunier2011,Russ2017a,Gungordu2018a}.
To this end, let us denote the four eigenvectors of $H|_{\Omega_i=0}$ as $\ket{\uparrow\uparrow},\ket{\psi_+(t)}, \ket{\psi_-(t)}, \ket{\downarrow\downarrow}$.
We then transform to the logical basis of $\{e^{i\phi_{\uparrow\uparrow}(t)}\ket{\uparrow\uparrow},e^{i\phi_+(t)}\ket{\psi_+(t)}, e^{i\phi_-(t)}\ket{\psi_-(t)}, e^{i\phi_{\downarrow\downarrow}(t)}\ket{\downarrow\downarrow}\}$ using $H_R = R^\dagger H R + i \hbar (\partial_t R^\dagger) R$, where $R$ is the transformation matrix whose rows are given by the logical basis states and $\phi_i(t)$ are yet undetermined degrees of freedom, associated with shifts in $ZZ$, $IZ$, $ZI$ terms in the Hamiltonian in this new frame ($ZZ$ denotes the tensor product of the Pauli matrices $\sigma_z$ and $\sigma_z$, $IZ$ denotes the tensor product of the identity and $\sigma_z$, and etc.). We thus obtain the rotating frame Hamiltonian
\begin{widetext}
\begin{align}
H_R = \begin{pmatrix}
\bar E_z + \dot \phi_{\uparrow\uparrow} & \frac{(\tilde E_{\perp}^{2,+})^*}{2} e^{-i(\phi_{\uparrow \uparrow} - \phi_+)} & \frac{(\tilde E_{\perp}^{1,+})^*}{2} e^{-i(\phi_{\uparrow \uparrow} - \phi_-)} & 0 \\
\frac{\tilde E_{\perp}^{2,+}}{2} e^{i(\phi_{\uparrow \uparrow} - \phi_+)} &  \frac{1}{2}(-J + \Delta E) + \dot\phi_+ & \tilde V & \frac{(\tilde E_{\perp}^{1,-})^*}{2} e^{i(\phi_{\downarrow \downarrow}-\phi_+)} \\
\frac{\tilde E_{\perp}^{1,+}}{2} e^{i(\phi_{\uparrow \uparrow} - \phi_-)} & \tilde V^* &  \frac{1}{2}(-J - \Delta E) + \dot\phi_-& \frac{(\tilde E_{\perp}^{2,-})^*}{2}e^{i(\phi_{\downarrow \downarrow}-\phi_-)} \\
0 & \frac{\tilde E_{\perp}^{1,-}}{2} e^{-i(\phi_{\downarrow \downarrow}-\phi_+)} & \frac{\tilde E_{\perp}^{2,-}}{2} e^{-i(\phi_{\downarrow \downarrow}-\phi_-)} & - \bar E_z + \dot \phi_{\downarrow\downarrow},
\end{pmatrix}
\end{align}
\end{widetext}
where $\Delta E = \sqrt{\delta E_z^2 + J^2}$ is the energy splitting between the adiabatic eigenstates of $H$ with zero net spin, $\tilde V$ is a diabatic correction which vanishes when $J$ and $\delta E_z$ do not vary in time \cite*{Gungordu2018a}. The transverse terms in the logical adiabatic basis are given by \cite*{Gungordu2018a}
\begin{align}
\tilde E_{\perp}^{1,\pm} = \tilde \Omega_1^\pm e^{i \omega t} = \frac{ (\Delta E + \delta E_z)\Omega_1 \mp J \Omega_2}{ \sqrt{2\Delta E (\Delta E + \delta E_z)}} e^{i \omega t}, \nonumber\\
 E_{\perp}^{2,\pm} = \tilde \Omega_2^\pm e^{i \omega t} = \frac{(\Delta E + \delta E_z) \Omega_2  \pm J \Omega_1 }{ \sqrt{2\Delta E (\Delta E + \delta E_z)} } e^{i \omega t}.
 \end{align}
This exchange-induced mixing of one-qubit ``Rabi frequencies", which we recognize as a form of crosstalk, is a result (and the cost) of working in the adiabatic basis with the simplified Ising-type exchange coupling.
We proceed by fixing the gauge with the choice $\dot\phi_\pm = \mp \Delta E/2$ and $\dot\phi_{\uparrow\uparrow} =\dot\phi_{\downarrow\downarrow}= -\bar E_z$, and tune the microwave frequency to $\hbar\omega = \bar E_z-\Delta E/2$ such that if the rotating wave approximation held and the ``fast" oscillating terms were negligible, we would have addressed only the second qubit (we can similarly tune into the first qubit by tuning the microwave as $\hbar\omega = \bar E_z+\Delta E/2$). We can write the resulting Hamiltonian
\begin{align}
H_R = \begin{pmatrix}
0 & \frac{\tilde\Omega_2^+}{2}  & \frac{\tilde\Omega_1^+}{2} e^{\frac{i}{\hbar} \Delta E t } & 0 \\
\frac{\tilde\Omega_2^+}{2} &  -\frac{J}{2} & 0 & \frac{\tilde\Omega_1^-}{2} e^{\frac{i}{\hbar}\Delta E t} \\
\frac{\tilde\Omega_1^+}{2} e^{-\frac{i}{\hbar} \Delta E t } & 0 &  -\frac{J}{2} & \frac{\tilde\Omega_2^-}{2} \\
0 & \frac{\tilde\Omega_1^-}{2} e^{-\frac{i}{\hbar}\Delta E t} & \frac{\tilde\Omega_2^-}{2}  & 0
\end{pmatrix}.
\end{align}
We note in passing that we will consider modulation of the envelope function $\Omega_i$ as a function of time as well in what follows, but these modulations will be much slower than $\omega$, so this decomposition of the microwave field into amplitude and frequency is still meaningful.

In the experiments in Refs. \cite*{Zajac2017a,Watson2017}, $\delta E_z$ is provided by an on-chip micromagnet which results in a strong magnetic field gradient, and thus a $\Delta E$ that is much larger than $\Omega_1$ such that a rotating wave approximation is valid. However, in the setups of Refs. \cite*{Veldhorst2014b,Huang2019}, $\delta E_z$ is due to the differences between $g$-factors (which can be modulated by the gate voltages \cite*{Chan2018a} or by changing the orientation of the external magnet \cite*{Tanttu2019}), and leads to a smaller $\delta E_z/h \sim 10$MHz, in which case the rotating wave approximation no longer holds.

Charge noise causes fluctuations in the exchange coupling $J \to \tilde J = J + \delta J$. Furthermore, spinful isotopes cause stochastic fluctuations in Zeeman splittings, $\varepsilon_1$ and $\varepsilon_2$, through hyperfine coupling.
With these sources of errors in mind, we can write the overall Hamiltonian as the sum of a noisy control and an oscillatory term, $\tilde H_R = \tilde H_c + H_\text{osc}$ where
\begin{align}
\label{eq:Hexplicit}
\tilde H_c =& \frac{\gamma \Omega_2}{2} IX + \frac{\eta \Omega_1}{2} ZX + \frac{\tilde J}{4} ZZ + \varepsilon_1 ZI + \varepsilon_2 IZ, \nonumber\\
H_\text{osc} =& \frac{\gamma \Omega_1}{2} [XI \cos(\Delta E t/\hbar) + YI \sin(\Delta E t/\hbar)] + \nonumber \\
&\frac{\eta \Omega_2}{2} [XZ \cos(\Delta E t/\hbar) + YZ \sin(\Delta E t/\hbar)]
\end{align}
and
\begin{align}
\gamma = \frac{(\Delta E + \delta E_z)}{ \sqrt{2\Delta E (\Delta E + \delta E_z)} }, \quad
\eta = \frac{J  }{ \sqrt{2\Delta E (\Delta E + \delta E_z)} } .
\end{align}
We will limit ourselves to exchange strengths smaller than the difference in Zeeman splittings, such that $J/\delta E_z$, despite being non-negligible \footnote{Although we are not aware of any such experiments, it is conceivable that in some devices, the minimal accessible value of the exchange in the (1,1) charge stability region may not allow treating $J_\text{min}/\delta E_z$ as a smallness parameter, in which case the first-order perturbation would not be sufficient and higher order corrections would need to be taken into account.}, can nevertheless be treated as smallness parameter allowing a perturbative treatment. Then, to the leading order in $J/\delta E_z$, $\gamma \approx 1$ and $\eta \approx J/2\delta E_z$.
Moreover, we can neglect the perturbations in $\gamma$ and $\eta$ due to exchange noise, because the $\delta J$ terms are suppressed by the smallness factor of $J/\delta E_z^2$ and we limit ourselves to $J \ll \delta E_z$. Exchange noise in $\Delta E$ can also be neglected where it is suppressed by the smallness factor $J/\delta E_z$.

At this point, we note that the control Hamiltonian $H_c$ fits into the embedding $\mathfrak{su}(2) \oplus \mathfrak{su}(2) \oplus \mathfrak u(1) \subset \mathfrak{su}(4)$. The  generators of these commuting $\mathfrak{su}(2)$ algebras are
 ${X_\pm, Y_\pm, Z_\pm} \equiv \{(IZ\pm ZZ)/2,  -(IY\pm ZY)/2,(IX\pm ZX)/2\}$ and the $\mathfrak u(1)$ is generated by $Q \equiv ZI$. Explicitly, $H_c$ can be split into three commuting parts
 \begin{align}
 H_c = H_+ + H_- + H_q
 \end{align}
 where
 \begin{align}
H_\pm \equiv& \Omega_\pm Z_\pm + \beta_\pm X_\pm   \qquad  H_q \equiv \varepsilon_1 Q \nonumber\\
\Omega_\pm \equiv&  \frac{\gamma\Omega_2}{2}\left(1 \pm \frac{1}{\gamma}\frac{g_1}{g_2}\eta \right) , \qquad \tilde\beta_\pm \equiv \varepsilon_2  \pm  \frac{\tilde J}{4},
\label{eq:su2su2u1}
 \end{align}
 where we used Eq.~\eqref{eq:g} to express $\Omega_1$ as $\Omega_2 g_1/g_2$.

The $\varepsilon_i$ terms can be made negligible by using a purified silicon with lower concentration of spinful silicon isotopes \cite*{Abrosimov2017}, which we will assume in what follows \footnote{We remark that robust pulses which correct for $\delta J$ also correct for $\varepsilon_2$ as well, since they combine together to form errors in $\beta_\pm$.}.

\section{Bloch-Siegert shift for the two-qubit system}
\label{sec:bloch-siegert}
In this section, we will apply the rotating wave approximation and also show how to compensate for the leading order corrections to the approximation, and in the following section, we will show how one can obtain a shaped pulse $\Omega_2(t)$ which can suppress the leading order effect of the terms $\delta J$ and $\eta$ from the final time evolution operator $U(t_f)$.

The validity of the rotating wave approximation hinges on the value of $\delta E_z/h$ (compared to the driving strength $\Omega_2/h$, as we will detail below).
This value is $1.3$GHz in the Si/SiGe device from Ref.~\cite*{Watson2017} and $210$MHz in the Si/SiGe device from Ref.~\cite*{Zajac2017a}, both of which incorporate micromagnets and operate via EDSR. However, in the Si/SiO$_2$ device from Ref.~\cite*{Huang2019,Yang2019b}, operating via ESR, this value is limited to around 15MHz, for which the rotating wave approximation to neglect $H_\text{osc}$ fails, resulting in poor gate fidelities (parameter values for this device are summarized in Table. \ref{tbl:values}).
In this section, we will show that the most significant contribution from the oscillatory terms can nevertheless be easily compensated.

To this end, we treat $H_\text{osc}$ as an interaction Hamiltonian and obtain the total time evolution operator as a product of the time evolution operators due to $H_c$ and $H_\text{osc}$ as
\begin{align}
U_\text{total} =  U_c(t_f;-t_f) U_\text{osc}(t_f;-t_f)
\label{eq:Utotal}
\end{align}
where $U_\text{osc}(t;-t_f)$ is the solution to the Schr\"odinger equation
\begin{align}
i \dot U_\text{osc}(t;-t_f) = \bar H_\text{osc}(t) U_\text{osc}(t;-t_f) \nonumber\\
\bar H_\text{osc} \equiv U_c^\dagger(t;-t_f) H_\text{osc}(t) U_c(t;-t_f)
\label{eq:Uosc}
\end{align}
When $\Delta E$ is larger than the microwave amplitude, one can obtain $U_\text{osc}$ in a perturbative manner using Magnus expansion
\begin{align}
U_\text{osc}(t;-t_f) = e^{\sum_{n=1}^\infty \Phi_n},
\end{align}
This allows use to obtain corrections to the rotating wave approximation perturbatively in powers of the smallness factor $\Omega_2(t)/\Delta E$, as will be made clear shortly.
The first order term in the Magnus expansion,
\begin{align}
\Phi_1 = -\frac{i}{\hbar} \int_{t_0}^{t_f} dt' \bar H_\text{osc}(t'),
\end{align}
is negligible when $\Delta E$ is sufficiently larger than the microwave amplitude.
For example, for $\text{max}|\gamma\Omega_2|=1$MHz, we find that a value of $\Delta E/h \sim 10$MHz, which is readily attainable in experiments, is sufficiently large.
We note that this condition is also relevant for the convergence of the Magnus expansion, which requires that $||\Phi_1|| < \pi$.

The leading order correction to the dynamics, then, is
\begin{align}
\Phi_2 = -\frac{1}{2\hbar^2} \int_{t_0}^{t_f} dt' \int_{t_0}^{t'} dt'' [\bar H_\text{osc}(t'), \bar H_\text{osc}(t'')].
\end{align}
We now show how this can be reduced to a simple condition with good accuracy.

The first simplification we make is to note that in $H_\text{osc}$, given explicitly in Eq.~\eqref{eq:Hexplicit}, the contribution of $\eta$ is suppressed by the smallness factor $J/\Delta E$ compared to $\gamma$. Thus, we neglect the terms proportional to $\eta$.
The ``fast" oscillations in $H_\text{osc}$ can be eliminated by (temporarily) going into a rotating frame defined by the transformation $S = e^{i \Delta E t ZI}$, which leads to
\begin{align}
H_\text{osc}^S = \frac{\gamma \Omega_2(t)}{2} XI + \Delta E \ ZI
\end{align}
Since $S \in \text{U}(1)$ commutes with $U(t;t_0) \in \text{SU}(2) \times \text{SU}(2)$, we can change the ordering of these transformations, and obtain
\begin{align}
U_\text{osc}(t;t_0) = e^{\frac{i}{\hbar} \Delta E t ZI} \mathcal T e^{-\frac{i}{\hbar}\int dt U_c^\dagger(t;t_0) H_\text{osc}^S(t) U_c(t;t_0)},
\end{align}
back in the frame of Eq.~\eqref{eq:Hexplicit}.
Although we can use second order Magnus expansion given by $\Phi_2$ as is to approximately evaluate the time-ordered integration above at this point, we instead notice that $ZI$ term in the integrand above commutes with $U(t;t_0)$, and the weak (when compared to $\Delta E$) term  $\gamma \Omega_2(t) U^\dagger(t;t_0) [XI] U(t;t_0)/2$ anticommutes with $ZI$ at all times, forming a dynamical $\mathfrak{su}(2)$ algebra. Based on this observation, we approximate the leading order correction $\Phi_2$ by an averaged Bloch-Siegert shift and obtain:
\begin{align}
U_\text{osc}(t_f;t_0) \approx \exp\left[{-i \frac{1}{\hbar \Delta E} \int_{t_0}^{t_f} dt \left(\frac{\gamma\Omega_2(t)}{2}\right)^2 ZI}\right].
\label{eq:BlochSiegert}
\end{align}
The inverse of this rotation can then be applied prior to the actual gate $U_\text{total}$ (cf. Eq.~\eqref{eq:Utotal}) to compensate for the unwanted $ZI$ Bloch-Siegert rotation.

An alternative way of obtaining this correction would be to solve the Schr\"odinger equation Eq.~\eqref{eq:Uosc} numerically. $U_\text{osc}$ tends to be a rotation mostly around $ZI$, so we can pick up the $ZI$ part of this SU(4) rotation, which can be corrected for easily, using
\begin{align}
U_\text{osc}^{(ZI)} = e^{i \Theta ZI}, \quad \Theta \equiv \text{tr}\left(-i \ln [U_\text{osc}(t_f;-t_f)] ZI\right).
\label{eq:BlochSiegert2}
\end{align}
We numerically find that the approximations we made in this section lead to deviations in trace gate fidelity which depend on $\Omega_2(t)/\Delta E$, and can be as small as  $\sim 10^{-4}$ with a suitable and experimentally attainable choice when using the parameters from Refs. \cite*{Huang2019,Yang2019b} (summarized in Table. \ref{tbl:values}). The numerical results for each gate are given in Section~\ref{sec:gates}.

Overall, the results from this section allow us to easily compensate for the shortcomings of the rotating wave approximation by applying a (virtual \cite*{McKay2017,Knill2008,Vandersypen2005,Huang2019}) $ZI$ rotation given in Eq.\eqref{eq:BlochSiegert} to undo $U_\text{osc}(t_f;t_0)$ in Eq.~\eqref{eq:Utotal}. This compensation can be done before, after or during the gate; the ordering does not matter because $ZI$ is the $\mathfrak u(1)$ generator and commutes with the control Hamiltonian.
We thus need not carry $H_\text{osc}$ forward in what follows, as it is now compensated for separately.

\section{Robust gates}
\label{sec:robustness}
In this section, we briefly summarize the results from \cite*{Barnes2015}, and show how they can be adapted to the double quantum dot system. In the next section, we will present specific applications for obtaining a robust CZ gate and robust, arbitrary one-qubit gates.

Consider a time evolution from $t=0$ to $t_f$ of a two-level system described by the noisy Hamiltonian
\begin{align}
H(t) = [\Omega(t)+ g(t)\epsilon_Z]Z + [\beta+\epsilon_X]X.
\end{align}
where $\epsilon_Z$ and $\epsilon_X$ are quasistatic, stochastic noise terms. The leading order effects of these noise terms from the time evolution at the final time $t_f$ can be eliminated if $\Omega(t)$ is a shaped pulsed obtained from a function $\Phi(\chi(t))$, through the following relation
\begin{align}
\label{eq:Omega}
\Omega(t) = \bar\Omega(\chi)   \equiv& -\beta \sin(2\chi) \times  \\
&\frac{\Phi''(\chi) + 4\Phi'(\chi)\cot(2\chi) + [\Phi'(\chi)]^3\sin(4\chi)}{ 2\sqrt{1 + [\Phi'(\chi)\sin(2\chi)]^2}^3 } \nonumber
\end{align}
where $\chi = \chi(t)$ is a reparametrization of time determined by $\Phi(\chi)$ through the relation
\begin{align}
\beta t = \hbar \int_0^{\chi_f} d\chi \sqrt{1 + [\Phi'(\chi)\sin(2\chi)]^2},
\label{eq:t}
\end{align}
$\Phi(\chi)$ is any function which obeys the constraints \cite*{Barnes2015}
\begin{subequations}
\begin{align}
\sin(4\chi_f) + 8 e^{ -2i \Phi(\chi_f)} \int_0^{\chi_f} d\chi \sin^2(2\chi) e^{ 2 i \Phi(\chi)} = 0,
\label{eq:robustness1}
\end{align}
\begin{align}
\int_0^{\chi_f} \sin^2(2\chi) \Phi'(\chi) = 0,
\label{eq:robustness2}
\end{align}
\begin{align}
\int_0^{\chi_f} d\chi \sin(2\chi) \bar g(\chi) e^{ 2 i \Phi(\chi)}  \sqrt{1 + [\Phi'(\chi)\sin(2\chi)]^2} = 0
\label{eq:robustness3}
\end{align}
\begin{align}
\int_0^{\chi_f} d\chi \cos(2\chi) \bar g(\chi)  \sqrt{1 + [\Phi'(\chi)\sin(2\chi)]^2} = 0,
\label{eq:robustness4}
\end{align}
\end{subequations}
and $\bar g(\chi) \equiv g(t)$.
The function $\Phi(\chi)$ must satisfy the following initial conditions
\begin{align}
\Phi(0) = 0, \qquad \Phi'(0) = 0
\label{eq:Phi0}
\end{align}
to ensure that the initial time evolution operator is the identity. The resulting gate $U(t_f;0)$ is determined by the values of $\Phi(\chi_f)$ and $\Phi'(\chi_f)$ \cite*{Barnes2015}.

We focus on the particular case of $g(t) = \Omega(t)$ (i.e., multiplicative noise in the control field), or equivalently, $\bar g(\chi) = \bar \Omega(\chi)$. If $\Phi'(\chi)$ is an odd function of $\chi$ (which itself is an odd function of $t$), the pulse shape $\bar\Omega(\chi)$ becomes an odd function of time as well.
Since this means the integrand of robustness conditions Eqs.~\eqref{eq:robustness2} and ~\eqref{eq:robustness4}
are odd functions for time, we can consider expanding the time evolution to the symmetric interval from $t=-t_f$ to $t=t_f$, which ensures that the integrals vanish and both conditions are satisfied \cite*{Barnes2015}. When the system is pulsed from $t=t_0=-t_f$ to $t=t_f$ using a pulse that is odd in time, the resulting time evolution is given by \cite*{Barnes2015,Gungordu2019b}
\begin{align}
\label{eq:U}
&U(t_f;-t_f) = Z_\phi X_\theta Z_{-\phi} = e^{-i\frac{\theta}{2}(\cos\phi X + \sin\phi Y)},\\
&\phi \equiv  \text{sgn}\left[\Phi'(\chi_f)\right]\text{arcsec}{ \sqrt{1 + [\Phi'(\chi_f)\sin(2\chi_f)]^2} }, \quad \theta \equiv 4\chi_f,\nonumber
\end{align}
where $X_\theta$ is a $\theta$ rotation around the $X$ axis and $Z_\phi$ is defined similarly. We refer to the Supplementary Information in Ref. \cite*{Barnes2015} for the lengthy derivation of these results.

We now show how these results can be used to suppress the effects of $\delta J$ and $\eta$ in $\tilde H_c$ from the final time evolution operator $U(t_f)$ in the double quantum dot system.
Although the four-level problem is algebraically split into a commuting pair of two-level systems and a single-level system, or $\mathfrak{su}(2) \oplus \mathfrak{su}(2) \oplus \mathfrak{u}(1)$,
we see from Eq.~\eqref{eq:su2su2u1} that the dynamics of these $\mathfrak{su}(2)$s are completely dependent: that is, choosing the pulse shape $\Omega_\pm(t)$ and the (time-independent) ``energy splitting" $\beta_\pm$ for one $\mathfrak{su}(2)$ fixes the values $\Omega_\mp(t)$ and $\beta_\mp$ for the other $\mathfrak{su}(2)$. This means that when looking for pulse shapes that can correct for errors, additional care is required when choosing suitable generating functions $\Phi_\pm(\chi)$ for each $\mathfrak{su}(2)$, and that $\Phi_+(\chi)$ and $\Phi_-(\chi)$ cannot be chosen independently ---in fact, choosing one completely determines the other.

We now show that even under this constraint, we can still use the results from Ref. \cite*{Barnes2015} to implement a robust quantum gate in the four-level system when the exchange is held fixed and the microwave source is smoothly pulsed in time.

First, from Eq.~\eqref{eq:su2su2u1}, we observe that in the absence of the stochastic noise and $\eta$ terms, we have
\begin{align}
\beta_+ = -\beta_-, \qquad \Omega_+(t) = \Omega_-(t).
\end{align}
Keeping Eq.~\eqref{eq:Omega} in mind, we realize that the choice
\begin{align}
\Phi_-(\chi) = -\Phi_+(\chi)
\end{align}
 would be compatible with these dynamical constraints between the $\mathfrak{su}(2)$ subsystems. This motivates us to treat the entire $\eta \Omega_2(t) ZX g_1/2 \gamma g_2 $ term as ``noise", and
 not just the (already neglected) truly stochastic part which is due to $\delta J$; this assumption is not strictly necessary, but it is convenient as it allows us to establish a very simple relationship between the two generating functions $\Phi_-(\chi)$ and $\Phi_+(\chi)$.

With this choice, the overall robust time evolution for the four-level system can thus be obtained as
\begin{align}
\mathcal R(\theta,\phi) \equiv& \mathcal R_+(\theta,\phi) \mathcal R_-(\theta,\phi),\nonumber\\
\mathcal R_\pm(\theta,\phi) \equiv& Z_\pm(\phi) X_\pm(\pm \theta) Z_\pm(-\phi),
\end{align}
where $\mathcal R_\pm$
denotes the time evolution for each $\mathfrak{su}(2)$ subsystem. The rotation angle $\theta$ has alternating signs because $\beta_+ = -\beta_-$, corresponding to time-inversion following Eq.~\eqref{eq:t}, which implies a  sign flip in the overall rotation angle in Eq.~\eqref{eq:U}. Combining similar commuting terms, this simplifies to
\begin{align}
\mathcal R(\theta,\phi) = IX_\phi ZZ_\theta IX_{-\phi}
\label{eq:Euler}
\end{align}
The pulse for the opposite sign, $\phi \to -\phi$, can be obtained by the replacement $\Phi(\chi) \to -\Phi(\chi)$ which implies $\Omega(t) \to -\Omega(t)$, following respectively  from Eqs.~\eqref{eq:U} and \eqref{eq:Omega}.
We will use this expression when targeting specific gates in the next section.

\section{Examples of robust gates}
\label{sec:gates}

For obtaining numerical results when targeting specific gates in this section, we will use the following ansatz:
\begin{align}
\Phi_+(\chi) = a_1 \chi^2 + \text{sgn}(\chi)\left[a_2 \chi^3 + \sum_{i=1}^8 b_i \sin(n \pi \chi/\chi_f)\right],
\end{align}
which obeys $\Phi_+(0) = \Phi_+'(0) = 0$ by construction. This ansatz leads to a smooth pulse shape $\Omega_+(t)$ that is odd in time, and readily satisfies the initial conditions given in Eq.~\eqref{eq:Phi0}. The coefficients $a_i$ and $b_i$ are free parameters which will be used to find suitable pulse shapes that implement a specific target unitary in the following sections. Furthermore, this form allows enforcing the rotation axis $\phi$ in Eq.~\eqref{eq:U} and the desirable property that the microwave source is turned off at the end, $\Omega_+(t_f)=0$, in an analytical manner:
\begin{align}
\label{eq:ai}
a_1 =& \frac{\tan\phi}{\chi_f \sin(2\chi_f)}\left(1+\chi_f \cot\left(2\chi_f\right) \left[1+\sec^2\phi\right] \right), \\
a_2 =& -\frac{\tan\phi}{3\chi_f^2 \sin(2\chi_f)}
\left(1+2\chi_f \cot\left(2\chi_f\right) \left[1+\sec^2\phi\right] \right) \nonumber
\end{align}
Finally, this leads to pulse shapes with modest bandwidth requirements $\Delta f \sim 4/t_f$.

We use the experimentally attainable values of $J/h=1$MHz and $\delta E_z/h = 15$MHz, which yields $\eta \approx 0.0333$ and $\gamma \approx 0.9994$. For a given $\phi$ and $\theta$, we use numerical constrained global optimization to solve for
the remaining robustness conditions Eqs.~\eqref{eq:robustness1} and ~\eqref{eq:robustness3}, subject to the constraints \footnote{We use the Improved Stochastic Ranking Evolution Strategy algorithm implementation in NLopt and limit the number of iterations to $10^5$. The numerical optimization takes around one minute on an Intel Westmere laptop CPU.}
\begin{align}
\label{eq:constraints}
\text{max } |\gamma \Omega_2(t)/h| \leq 1\text{MHz}, \qquad T \equiv 2t_f \leq 20\mu\text{s},
\end{align}
where $T$ is the total gate time. The first constraint is necessary for improving the accuracy of the Bloch-Siegert shift compensation, where we assumed that $\gamma\Omega_2(t)/\delta E_z$ is a smallness factor whereas the second condition is for keeping the gate time within reasonable limits.

\begin{table}
\begin{ruledtabular}
\begin{tabular}{|l|l|}
Parameter & Value \\
\hline
$\delta E_z$ & 15MHz\\
$J$ & $1$MHz \\
$\text{max}|\gamma \Omega_2(t)|$ & 1MHz or 1.5MHz \\
$\gamma$ & 0.9994\\
$\eta$ & 0.0333 \\
\end{tabular}
\end{ruledtabular}
\caption{Values of parameters used for examples, based on the Si/SiO$_2$ device from \cite*{Huang2019,Yang2019b}.}
\label{tbl:values}
\end{table}

We are only able to find numerical solutions within these constraints if we introduce additional windings to the $ZZ$ rotation as $\theta \to 2\pi k + \theta$. For a given target angle $\phi$, the gate time increases with $k$, which puts a limit on how small $\Omega_2(t)$ can be; we choose $k$ such that the gate time is minimal.

\subsection{Robust CZ gate}
\begin{figure}
	\includegraphics[width=1\columnwidth]{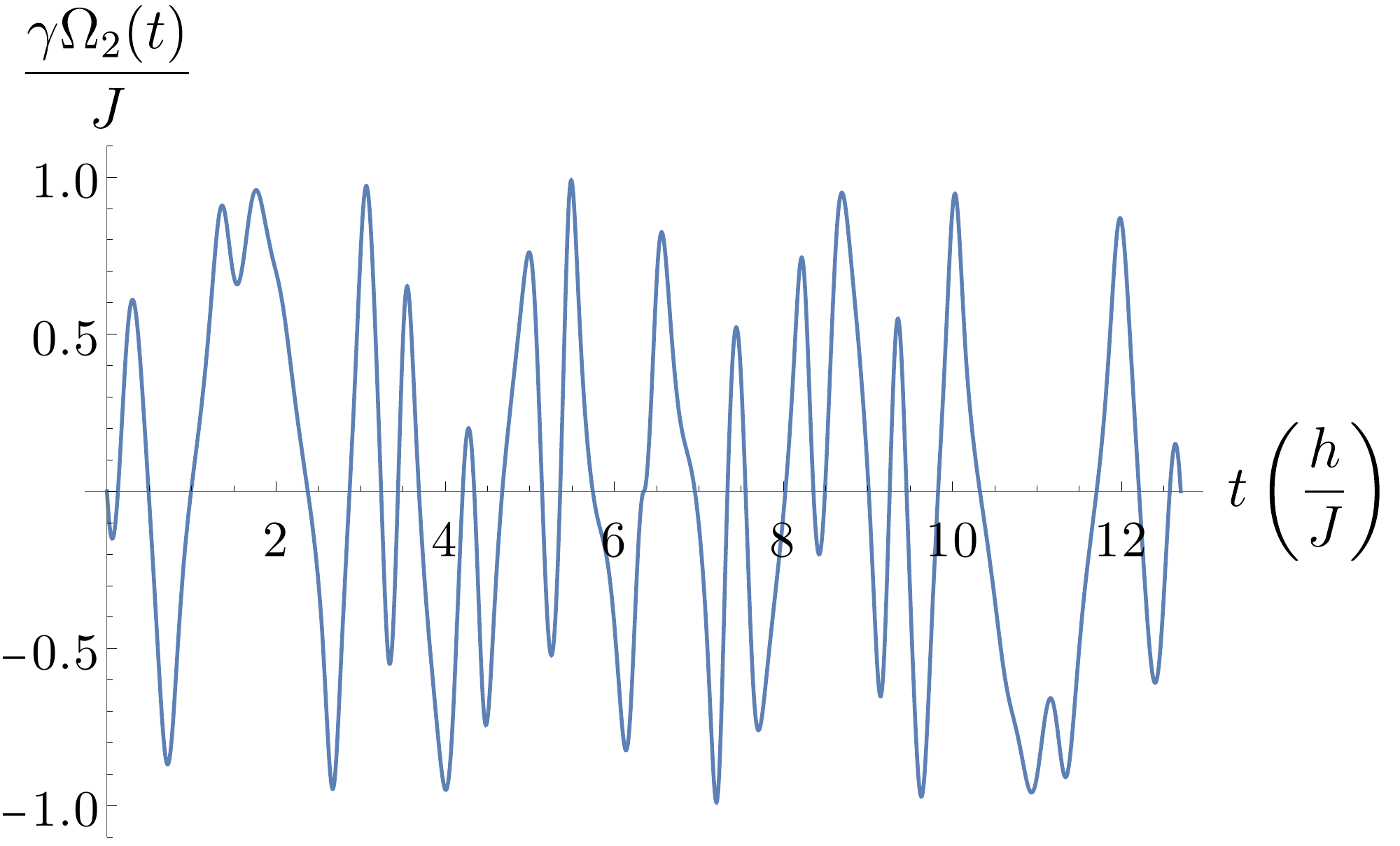}
	\includegraphics[width=1\columnwidth]{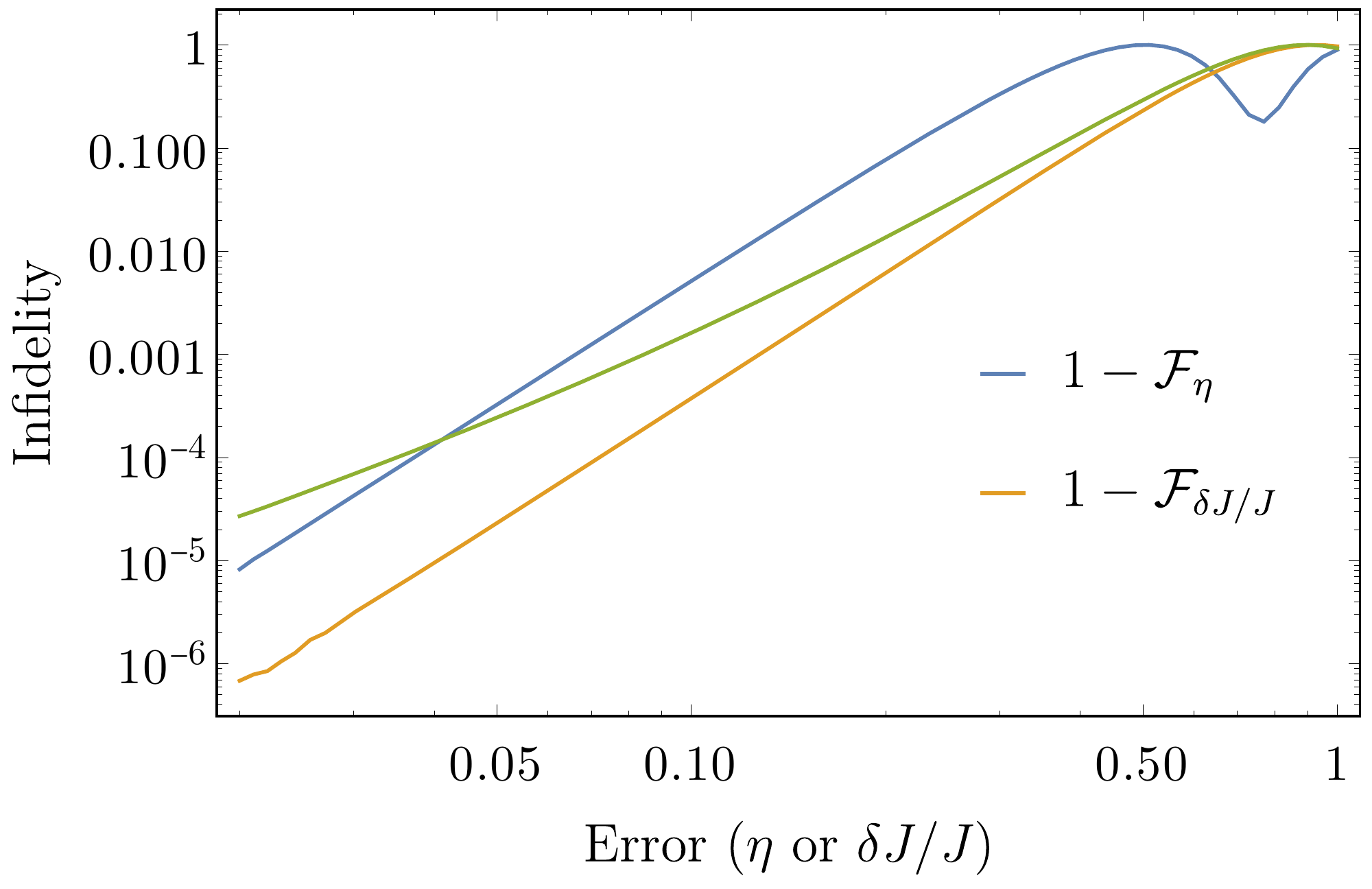}
	\caption{(Color online) (a) Pulse shape $\gamma\Omega_2(t)$ in units of $J=h \times 1$MHz which implements a CZ gate, $\mathcal R(\pi/2,0)$, in $T \approx 13\mu$s that is robust against exchange-induced crosstalk $ZX$ term whose strength is characterized by $\eta$, and perturbations in exchange $\delta J$. The pulse shape is determined by 8 parameters $b_i$ given in Eq.~\eqref{eq:biCZ}. (b) Infidelity of the robust pulse as a function of perturbation strength, $\eta$ or $\delta J/J$, compared to the infidelity of a naive pulse (green curve) with $\Omega_2=0$ as a function of $\delta J/J$.}
	\label{fig:CZ}
\end{figure}

A CZ gate $\exp\left({-i \frac{\pi}{4}ZZ}\right)$ corresponds to $\theta = \pi/2$ and and $\phi=0$. We take the target angle to be $\theta = 2\pi k + \pi/2$ with $k=5$, which determines the values of $a_i$ through Eq.~\eqref{eq:ai} as $a_1=0$ and $a_2=0$, and using numerical optimization for robustness conditions we find
\begin{align}
\boldsymbol{b} \approx \{ & -2.63, 0.07, -0.31, -0.52, 0.09, 0.01, -0.03, -0.06\}.
\label{eq:biCZ}
\end{align}
The resulting pulse shape $\Omega_2(t)$, which takes $\approx 12.7\mu$s is shown in Fig.~\ref{fig:CZ}.

Once the $ZI$ Bloch-Siegert shift Eq.~\eqref{eq:BlochSiegert} is compensated for using a virtual-$Z$ rotation, the infidelity due to neglecting $H_\text{osc}$ is $\approx 4 \times 10^{-4}$, whereas using Eq.~\eqref{eq:BlochSiegert2} yields $\approx 2 \times 10^{-4}$.
From Fig.~\ref{fig:CZ}, we see that an infidelity budget of $10^{-4}$ is able to tolerate errors up to $\delta J/J \approx 0.075$ (i.e., $\delta J/h \approx 75$kHz) and $\eta \approx 0.04$.

Implementing such a smooth pulse shape exactly may be challenging. As an example, we numerically checked that at $\delta J/h \approx 75$kHz exchange error, an imperfect pulse shape with  0.01 errors in all pulse ``amplitudes" $b_i$ results in $\approx 3 \times 10^{-4}$ infidelity instead of $\approx 10^{-4}$ with a perfect pulse shape.

We numerically find that the limit on the maximum allowed value of $\gamma\Omega_2/h$ can be raised to 1.5MHz with similar error characteristics, with a shorter gate time of $T \approx 9.2\mu$s at $k=3$, with parameters
\begin{align}
\boldsymbol b \approx \{-0.51, 1.62, 0.12, -0.20, 0.05, 0.09, 0.04, -0.01\}.
\end{align}

\subsection{Robust one-qubit gates}

\begin{figure}
	\includegraphics[width=1\columnwidth]{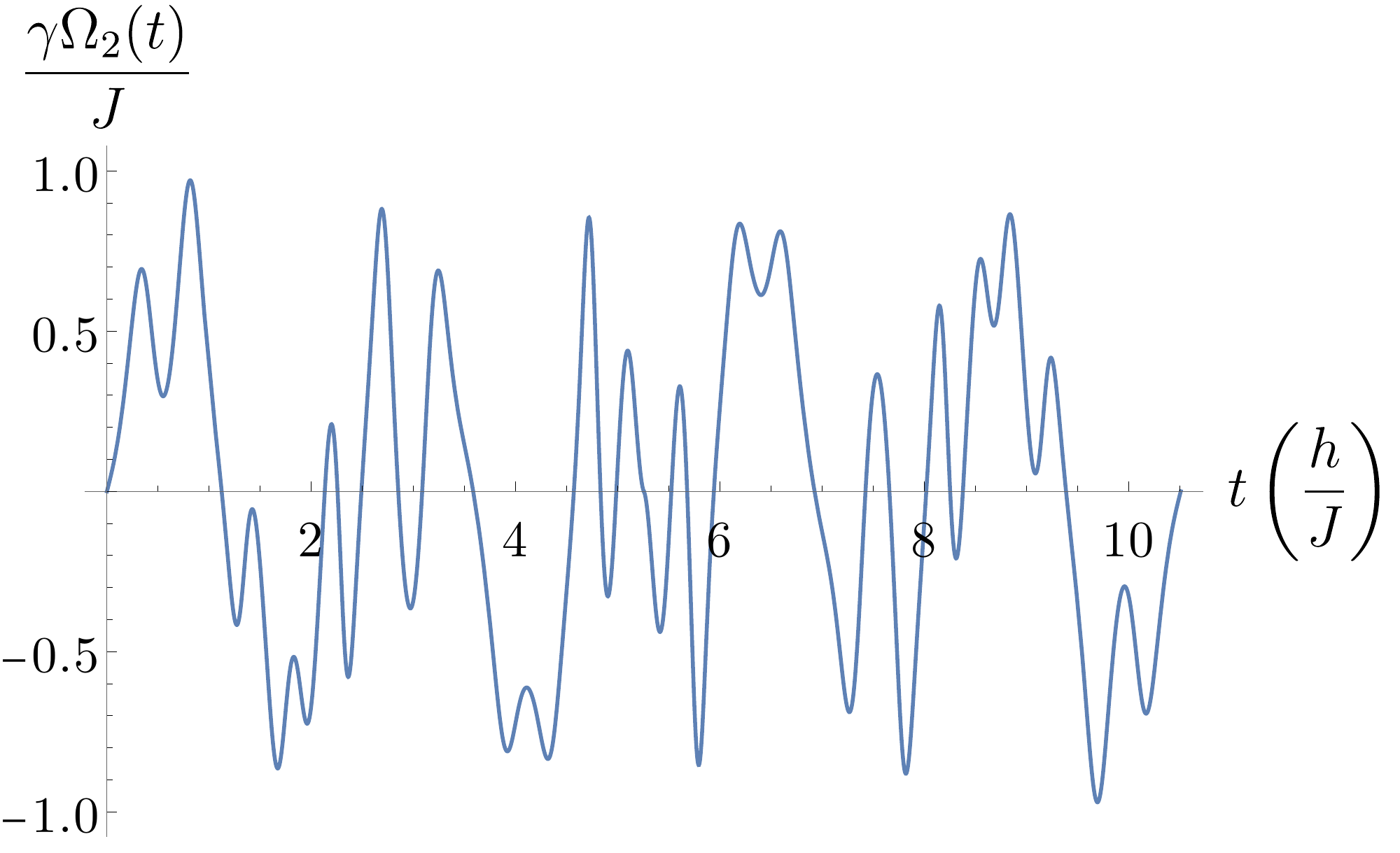}
	\includegraphics[width=1\columnwidth]{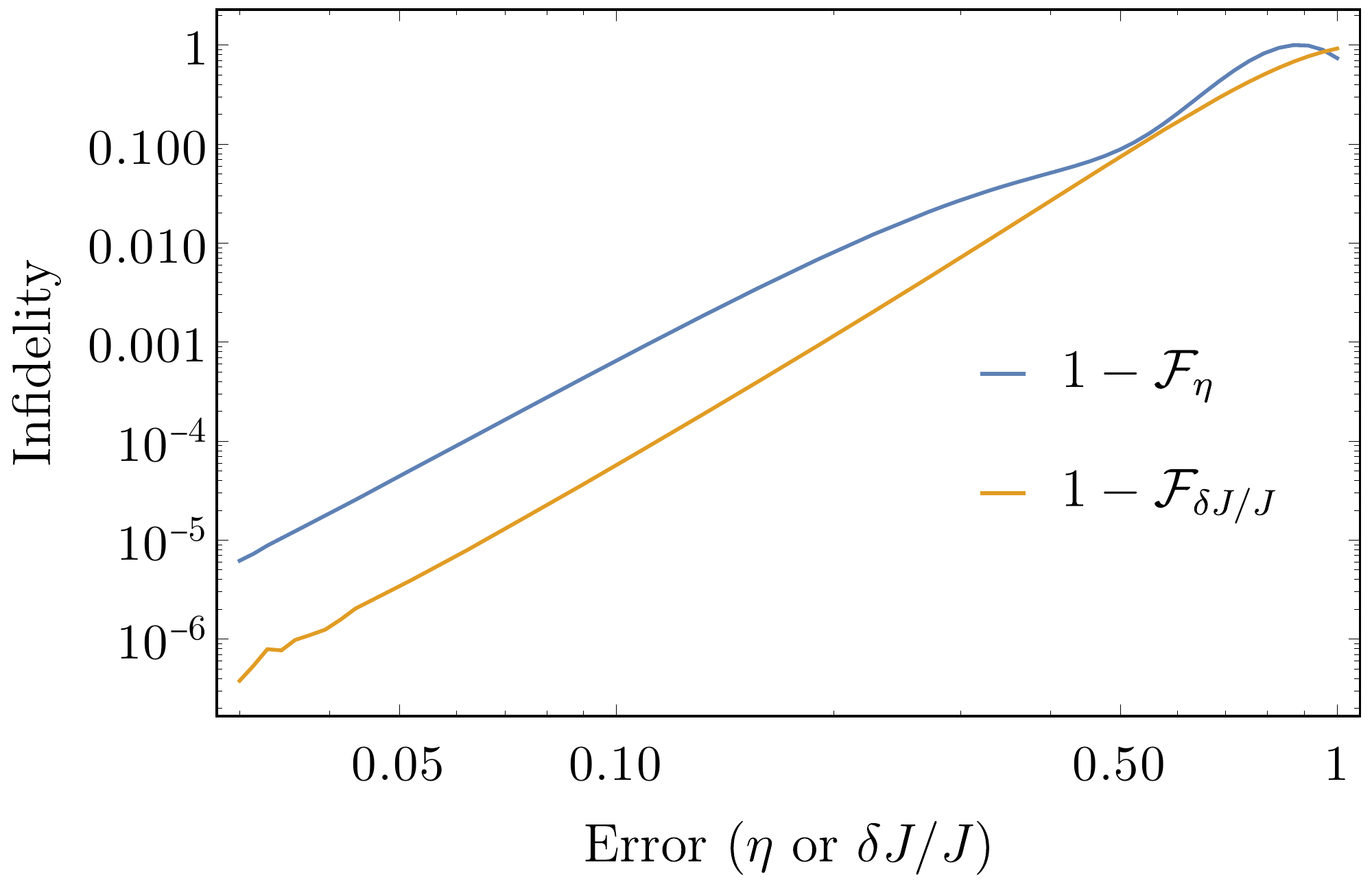}
	\caption{(Color online) (a) Pulse shape $\gamma\Omega_2(t)$ in units of $J=h \times 1$MHz which implements $\mathcal R(\pi,\pi/4)$, which is $IX_{\frac{\pi}{2}}$ up to local $IZ$ $\pi$-pulses, in $T \approx 11\mu$s that is robust against exchange-induced crosstalk $ZX$ term whose strength is characterized by $\eta$, and perturbations in exchange $\delta J$. The pulse shape is determined by 8 parameters $b_i$ given in Eq.~\eqref{eq:biXpi2}. (b) Infidelity as a function of perturbation strength, $\eta$ or $\delta J/J$.}
	\label{fig:Xpi2}
\end{figure}

For implementing one-qubit gates, we make use of the availability of virtual local $Z$ rotations \cite*{McKay2017,Knill2008,Vandersypen2005} (implemented by switching the phase of the microwave source instantaneously \cite*{Huang2019}) as follows. Any one-qubit gate can be expressed in terms of three Euler angles
\begin{align}
\label{eq:ZYZ}
IZ_{\alpha_3} IY_{\alpha_2} IZ_{\alpha_1} =& IZ_{\alpha_3} IX_{\frac{\pi}{2}} IZ_{\alpha_2} IX_{-\frac{\pi}{2}} IZ_{\alpha_1}\\
=& -IZ_{\alpha_3} IX_{\frac{\pi}{2}} IZ_{\alpha_2+\pi} IX_{\frac{\pi}{2}} IZ_{\alpha_1+\pi}. \nonumber
\end{align}
This implies that having access to a single robust one-qubit gate, $IX_{\frac{\pi}{2}}$, rather than a set of gates with a continuous parameter, is sufficient.

Such a robust one-qubit gate can be implemented using $\mathcal R(\theta,\phi)$ given in Eq.~\eqref{eq:Euler}, since at $\theta=\pi$ it reduces to the local gate
\begin{align}
\mathcal R(\pi,\phi) = IX_\phi ZZ_\pi IX_{-\phi} = IX_{2\phi} ZZ_\pi = -i IX_{2\phi} IZ_\pi  ZI_\pi
\end{align}
The extra $IZ$ and $ZI$ rotations can be cancelled using local virtual $Z$ gates which leaves a pure $IX$ rotation. In the context of Eq.~\eqref{eq:ZYZ}, however, this is unnecessary as they combine and cancel to give the relation
\begin{align}
IZ_{\alpha_3} IY_{\alpha_2} IZ_{\alpha_1} = -IZ_{\alpha_3} \mathcal R\left(\pi,\frac{\pi}{4}\right) IZ_{\alpha_2} \mathcal R\left(\pi,\frac{\pi}{4}\right) IZ_{\alpha_1}.
\label{eq:sequence3}
\end{align}
We can thus implement any arbitrary one-qubit gate robustly using $\mathcal R(\pi,\pi/4)$ in conjunction with virtual $IZ$ rotations.

For finding a pulse shape which implements $\mathcal R(\pi,\pi/4)$, we take $\theta = 2\pi k + \pi$ with $k=4$, and obtain the parameters $b_i$
\begin{align}
\label{eq:biXpi2}
\boldsymbol b \approx \{ & -2.52, 0.24, -0.24, -0.48, 0.21, 0.04, 0.02, -0.02\}.
\end{align}
The corresponding pulse shape is given in Fig.~\ref{fig:Xpi2}, which takes $\approx 10.5\mu$s. This implies that arbitrary one-qubit gates can be implemented robustly in $\approx 21\mu$s, when using the sequence Eq.~\eqref{eq:sequence3}.

With the $ZI$ Bloch-Siegert correction from Eq.~\eqref{eq:BlochSiegert}, the infidelity due to neglecting $H_\text{osc}$ is $\approx 3 \times 10^{-4}$, whereas using  Eq.~\eqref{eq:BlochSiegert2} yields $\approx 2 \times 10^{-4}$.
From Fig.~\ref{fig:Xpi2}, we see that an infidelity budget of $10^{-4}$ allows $\delta J/h \approx 110$kHz and $\eta \approx 0.06$.
Fig.~\ref{fig:Xpi2} does not include an infidelity curve for a naive implementation, because one-qubit gates with always-on $J$ coupling is a nontrivial problem even without any robustness requirements \cite*{Huang2019,Li2008}.
We also observe that at $\delta J/h \approx 110$kHz exchange error, an imperfect pulse shape with  0.01 deviations in all  $b_i$ increases the infidelity to $\approx 2 \times 10^{-4}$.

As in the case of CZ gate, we find that raising the limit on the maximum allowed value of $\gamma\Omega_2/h$ to 1.5MHz yields an $IX_{\pi/2}$ gate with similar error characteristics, and takes $T \approx 8.22\mu$s at $k=3$, with parameters
\begin{align}
\boldsymbol b \approx \{-0.66, -0.16, -0.22, 0.32, -0.02, 0.03, 0.00, -0.03\}.
\end{align}

We note that the gate time for one qubit gates can be improved by making a look up table for $\boldsymbol b$ corresponding to $\mathcal R(\pi,\phi)$ for each $2\phi \in [-\pi/2,\pi/2]$, and use instead the $ZXZ$ parametrization for one-qubit rotations as $IZ_{\alpha_3} IX_{\alpha_2} IZ_{\alpha_1}$, which would allow faster implementation of arbitrary one-qubit gates. Such a table with the granularity of $\pi/64$ in $\phi$ is given in Table \ref{tbl:one-qubit-b} in Appendix \ref{sec:one-qubit-b}.

Our one-qubit gate times for $X_\theta$ rotations are similar to the $8\mu$s duration of robust pulses obtained by GRAPE in Ref. \cite*{Yang2019b} in the same device, albeit in the (1,0) charge regime with no exchange or crosstalk.

\section{Summary and Conclusion}
\label{sec:conclusion}
Spin qubits in silicon quantum dots are a promising platform for realization of a fault-tolerant quantum computer due to existing fabrication techniques, spinless nature of the $^{28}$Si nuclei and electrical controllability of the electron wavefunctions. For single quantum dots, single-qubit fidelities are approaching to fault-tolerance thresholds. However, a useful quantum computer requires multiple qubits, which introduce new issues such as addressability and limited control over inter-dot interactions. Although a slanted micromagnet can be used to enhance the addressability, not having an on-chip micromagnet has the advantage of simplifying the device design, which is especially important in the context of scalability. When combined with an always-on exchange coupling between the electrons, implementing high-fidelity quantum gates even in absence of any noise becomes a nontrivial problem.

We have shown that despite the addressability and always-on exchange coupling in a silicon double quantum dot setup, it is possible to implement robust CZ gate and robust arbitrary one-qubit gates, which form a universal set. We estimate that a $\approx 10\%$ error in the exchange coupling at $J=1$MHz would lead to $\approx 99.97\%$ fidelity for the CZ gate and $\approx 99.98\%$ fidelity for one-qubit $X$ gates. The resulting pulse shapes require a modest bandwidth around $\sim 1$MHz,
and can lead to gates with high fidelity even with systematic errors caused by the signal generator.
We expect these pulses will improve the gate fidelities over their naive non-robust counterparts, despite the increased gate times.
The relevant decoherence timescale for a naive gate is $T_2^*$, whereas for a dynamically corrected pulse it is essentially $T_2$.
In silicon quantum dots, $T_2^*$ times are several milliseconds \cite*{Yoneda2017b} whereas $T_2^*$ times are several tens of microseconds \cite*{Yoneda2017b,Yang2019b}, and an overall rough estimate for the improvement in gate infidelities can be made by comparing $\sim (T/T_2)^2$ to $\sim (T_\text{naive}/T_2^*)^2$, which yields an improvement by a factor of $\sim25$ for the CZ gate, and $\sim1000$ for one-qubit gates. The  speed of the robust gates is mainly limited by $\delta E_z$, which prevents us from using a larger microwave amplitude and exchange strength typically accessible in these devices. In devices with weaker $\delta E_z$, such as \cite*{Petit2019} with $\delta E_z \approx 9$MHz, this further limits the suitable range of $J$ and $\Omega_2$. We note, however, that a $\delta E_z$ around 40MHz is attainable with $g$-factor modulation in Si/SiO$_2$ \cite*{Veldhorst2014b} which would improve the speed of the gates by a factor of $\approx 2.5$.

We expect our results can be readily implemented in existing devices without modifications.

\begin{table*}[t]
\begin{ruledtabular}
\begin{tabular}{|l|l|l|l|}
$2\phi$ & $T (\mu\text{s})$ & $k$ & $\boldsymbol b$ \\
\hline
$\frac{\pi}{32}$ & 14.97 & 6 & $\{-2.52, 0.24, -0.24, -0.48, 0.21, 0.04, 0.02, -0.02\}$ \\
$\frac{2\pi}{32}$ & 15.05 & 6 & $\{3.06, 0.01,   0.53, 0.45, -0.09, 0.00, -0.09, -0.04\}$ \\
$\frac{3\pi}{32}$ & 14.97 & 5 & $\{2.60, -0.34, 0.20,   0.35, -0.03, -0.07, -0.06, 0.12\}$ \\
$\frac{4\pi}{32}$ & 13.02 & 5 & $\{0.12, 1.07, 0.07, 0.03, 0.03,   0.11, -0.04, -0.04\}$ \\
$\frac{5\pi}{32}$ & 11.86 & 4 & $\{0.79, -0.97, 0.16,   0.13, -0.02, -0.10, -0.05, -0.01\}$ \\
$\frac{6\pi}{32}$ & 9.92 & 4 & $\{0.72, -0.87, 0.21, 0.17, 0.02, -0.09, -0.02, 0.01\}$ \\
$\frac{7\pi}{32}$ & 9.90 & 4 & $\{0.56, -0.72, 0.14, 0.14, -0.01, -0.05, 0.00, 0.06\}$ \\
$\frac{8\pi}{32}$ & 7.89 & 3 & $\{-0.11, 0.78, 0.04, 0.01, 0.05, 0.02, 0.04, 0.03\}$ \\
$\frac{9\pi}{32}$ & 7.75 & 3 & $\{0.01, 0.61, 0.10, 0.05, 0.06, 0.01, 0.01, 0.01\}$ \\
$\frac{10\pi}{32}$ & 7.49 & 3 & $\{-0.23, 0.49, -0.02, 0.03, 0.07, 0.06, 0.01, 0.03\}$ \\
$\frac{11\pi}{32}$ & 7.56 & 3 & $\{-0.26, 0.35, 0.00, -0.03, 0.09, -0.01, 0.03, 0.02\}$ \\
$\frac{12\pi}{32}$ & 7.57 & 3 & $\{-0.29, 0.21, -0.06, 0.13, 0.08, -0.04, 0.00, 0.01\}$ \\
$\frac{13\pi}{32}$ & 10.00 & 4 & $\{0.80, 0.27, 0.22, 0.19, 0.00, -0.07, 0.00, 0.03\}$ \\
$\frac{14\pi}{32}$ & 10.01 & 4 & $\{0.67, 0.30, 0.19, -0.15, -0.01, -0.01, 0.01, 0.02\}$ \\
$\frac{15\pi}{32}$ & 10.24 & 4 & $\{0.96, 0.47, 0.30, 0.01, 0.01, 0.09, 0.04, 0.04\}$ \\
$\frac{16\pi}{32}$ & 10.51 & 4 & $\{0.80, 0.30, 0.26, -0.22, -0.07, 0.07, 0.05, 0.03\}$
\end{tabular}
\end{ruledtabular}
\caption{A look-up table of $\boldsymbol b$ values for implementing a robust $IX_{2\phi}$ gate using $k$ additional $2\pi$ rotations in $\theta$, obtained numerically under the constraint that $\gamma \Omega_2(t)/h$ never exceeds 1MHz. The gate $X_{-2\phi}$ can be implemented with the replacement $a_i,b_i \to -a_i,-b_i$.}
\label{tbl:one-qubit-b}
\end{table*}

\begin{acknowledgements}
This research was sponsored by the Army Research Office (ARO), and was accomplished under Grant Number W911NF-17-1-0287.
\end{acknowledgements}

\appendix

\section{Parameters for one-qubit gates $X_{2\phi}$}
\label{sec:one-qubit-b}
The parameters for one-qubit gates are listed in Table \ref{tbl:one-qubit-b}.

\bibliography{siqd,extra}

\end{document}